\documentclass{llncs}
\usepackage{makeidx}  
\usepackage{graphicx}        
\usepackage{fancyvrb}
\usepackage{fixltx2e}
\usepackage[english,polish, german]{babel}
\usepackage{fancyhdr}
\usepackage{hyperref} 
\begin{document}
\pagestyle{fancy}
\lhead{Uncertainty-dependent data collection in vehicular sensor networks}
\chead{}
\rhead{}
\lfoot{The final publication is available at \href{http://www.springerlink.com/content/k555h222n056n126/}{www.springerlink.com}}
\cfoot{}
\rfoot{}
\title{Uncertainty-dependent data collection in vehicular sensor networks}
\titlerunning{Uncertainty-dependent data collection in vehicular sensor networks}  
%
\selectlanguage{polish}
\author{Bart"lomiej P"laczek}
\authorrunning{Bart"lomiej P"laczek} 
%
\tocauthor{Bart"lomiej P"laczek}
\institute{Faculty of Transport, Silesian University of Technology,\\Krasi"nskiego 8, 40-019 Katowice, Poland\\
\email{bartlomiej.placzek@polsl.pl}}

\maketitle              

\selectlanguage{english}

\begin{abstract}
Vehicular sensor networks (VSNs) are built on top of vehicular ad-hoc networks (VANETs) by equipping vehicles with sensing devices. These new technologies create a huge opportunity to extend the sensing capabilities of the existing road traffic control systems and improve their performance. Efficient utilisation of wireless communication channel is one of the basic issues in the vehicular networks development. This paper presents and evaluates data collection algorithms that use uncertainty estimates to reduce data transmission in a VSN-based road traffic control system. 
\end{abstract}
%
\section{Introduction}
Vehicular sensor network (VSN) combines wireless communication provided by vehicular ad-hoc network (VANET) with sensing devices installed in vehicles. Sensors available in vehicles gather data sets including localisations, speeds, directions, accelerations, etc. Thus, the vehicles participating in VSN can be used as the sources of information to determine accurately the traffic flow characteristics.

Monitoring of road and traffic conditions becomes an important application area of VSNs. This new technology creates a huge opportunity to extend the road-side sensor infrastructure of the existing traffic control, management and safety systems \cite{bplaczek:bib1,bplaczek:bib18}. A major drawback of the current-generation traffic monitoring systems is a narrow coverage due to high installation and maintenance costs. It is expected that the VSNs will help to overcome these limitations.

Unlike traditional wireless sensor networks, VSNs are not subject to major memory, processing, storage, and energy limitations. However, in dense urban road networks, where number of vehicles uses the same transmission medium for many purposes, the periodic transmissions of all sensed data may consume the entire channel bandwidth resulting in excessive congestion and delays in the communication network. These effects are a major impediment for the time-constrained control tasks and safety related services. Therefore the efficient use of the wireless communication channel is one of the basic issues in VSNs applications development \cite{bplaczek:bib1}.

This paper presents and evaluates three data collection algorithms that use uncertainty estimates to reduce the data transmission in a VSN-based road traffic control system. The uncertainty-dependent data collection algorithms were inspired by an observation that for many cases the scope of real-time vehicular data potentially available in VSN exceeds the needs of particular traffic control tasks. The underlying idea is to detect the necessity of data transfers on the basis of uncertainty evaluation. The advantage of the introduced approach is that it uses selective on-time queries instead of periodical data sampling. 

The rest of the paper is organised as follows. Related works are reviewed in Section 2. Section 3 describes the VSN-based road traffic control system. Algorithms for traffic data collection are presented in Section 4. Section 5 contains the results of an experimental study on data collection for the traffic control in a road network. Finally, in Section 6, conclusion is given.
%
\section{Related works}
The emergence of VSNs technologies has made it possible to use novel, more effective techniques to deal with the problems of road traffic control. Several traffic control algorithms have been developed in this field for signalised intersections. These adaptive signal control schemes use real-time sensor data collected from vehicles (e.g. their positions and speeds) to minimise travel time and delay experienced by drivers at road intersections. Most methods are based on wireless communication between vehicles and road-side control nodes (e.g. \cite{bplaczek:bib2,bplaczek:bib3}). In few proposed solutions, vehicle-to-vehicle communication is used for implementing the traffic control \cite{bplaczek:bib4,bplaczek:bib5}.

In the above cited studies, the real-time sensed data are assumed to be collected continuously from all vehicles in a certain area. Such periodical data sampling scheme may cause excessive congestion and latency in the communication network due to the bandwidth-limited wireless medium. Therefore, more research is needed to determine required input data sets as well as sampling rates that are necessary for the traffic control.

In the literature several methods have been introduced for wireless sensor networks that enable the optimisation of data collection procedures. Suppression based techniques have been demonstrated to be useful in reducing the amount of sensor data transmitted for monitoring physical phenomena \cite{bplaczek:bib6}. Temporal suppression is the most basic method: sensor readings are transmitted only from those nodes where a change occurred since the last transmission \cite{bplaczek:bib7}. Spatial suppression methods aim to reduce redundant transmissions by exploiting the spatial correlation of sensor readings \cite{bplaczek:bib8}. If the sensor readings of neighbouring sensor nodes are the same or similar, the transmission of those sensed values can be suppressed. Model-based suppression methods use divergence between actual measurements and model predictions to detect the necessity of data transfers \cite{bplaczek:bib9}. Implementing this approach requires a pair of dynamic models of the monitored phenomenon, with one copy distributed in the sensor network and the other at a base station. 

Another effective approach to the optimisation problem of data collection in sensor networks is the model-based querying approach, in which the sensor data are complemented by a probabilistic model of the underlying system \cite{bplaczek:bib10}. According to this methodology, sensors are used to acquire data only when the model is not sufficiently rich to answer the query with an acceptable confidence. Each query has to include user-defined error tolerances and target confidence bounds that specify how much uncertainty is acceptable in the answer.

In \cite{bplaczek:bib11} an uncertainty-dependent data collection method was proposed for the VSN-based traffic control systems. In this method, the necessity of data transfers is detected by uncertainty evaluation of traffic control decisions. The sensor data are transmitted from vehicles to the control node only at selected time moments. For the remaining periods of time the sensor readings are replaced by results of an on-line traffic simulation. The effectiveness of this method was verified in an experimental study on traffic control at isolated intersection.

\section{VSN-based road traffic control}
The purpose of the VSN-based traffic control system is to manage the traffic flow by controlling traffic signals (Fig. 1). VSN senses positions and velocities of vehicles in a road network. The control loop includes data collection module, which sends selective on-time queries to retrieve necessary traffic data from the VSN. At each time step, the set of data that has to be acquired is determined taking into account the uncertainty estimated during decision-making procedure. Traffic model is an important component which uses the acquired data for approximation of the current traffic state as well as prediction of its future evolution. A task of the decision making module is to select an optimal control action on the basis of the information delivered by traffic model, according to the control strategy.
%
%
\begin{figure}
\centering
\includegraphics [height=2.22cm] {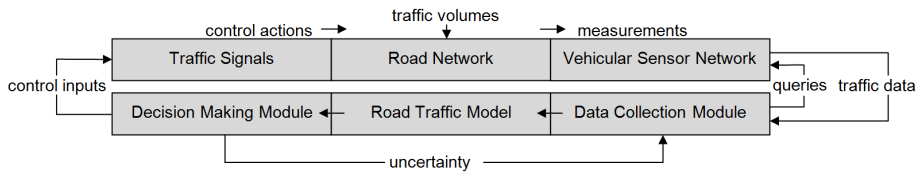}
\caption{VSN-based road traffic control system}
\end{figure}
%
\subsection{Traffic control strategy}
In the presented study, a decentralised self-control strategy \cite{bplaczek:bib12} was applied to minimise travel times in a road network. The self-organised traffic control is based on an optimisation and a stabilisation rule. Both rules are executed in parallel for all intersections in the network in order to adapt the traffic control to local flow conditions.

According to the self-organised traffic control strategy the consecutive control decisions are made in time steps of one second. A particular control decision determines which traffic stream should get a green signal at an intersection. The decision is made using the following formula:  
\begin{equation}
  \sigma = \left\{ 
  \begin{array}{l l}
     \mathrm{head} \: \mathrm{\Omega} & \quad \mathrm{if} \: \mathrm{\Omega} \neq \emptyset\\
    \mathrm{arg\:max}_i \pi_i & \quad \mathrm{otherwise,}\\
   \end{array} \right.
\end{equation}
where: $\sigma$ indicates the traffic stream which will get green signal, $\mathrm{\Omega}$ is an ordered set containing indices of the traffic streams that have been selected using the stabilisation rule, $\pi_i$ denotes priority of stream $i$, which is calculated on the basis of the optimisation rule.

The aim of stabilisation rule is to assure that all traffic streams will be served at least once in $T_{max}$ period. To this end, for each traffic stream a service interval $Z_i$ is predicted as the sum of preceding red time $r_i$ for stream $i$, intergreen time $\tau^0_i$ before switching the green signal for stream $i$, and green time $G_i$ required for vehicles in lane $i$ to pass the intersection:
\begin{equation}
  Z_i=r_i+\tau^0_i+G_i.
\end{equation}
The index $i$ of traffic stream joins the set $\mathrm{\Omega}$  as soon as $Z_i \geq T_{max}$.

\selectlanguage{german}
Optimisation rule aims for minimising waiting times by serving the incoming traffic as quickly as possible. According to this rule a traffic stream with the highest priority index $\pi_i $ gets green signal, provided that the set $\mathrm{\Omega}$ is empty. The priority index for stream $i$ is defined as
\begin{equation}
  \pi_i =\frac{N_i}{\tau^\mathrm{pen}_{i,\sigma}+\tau_i+G_i},
\end{equation}
where: $N_i$ denotes number of vehicles in lane $i$ that are expected to pass the intersection in time $\tau_i+G_i$, $\tau^\mathrm{pen}_{i,\sigma}$ is a penalty for switching from stream $\sigma$  to $i$, $\tau_i$ denotes intergreen time after green signal for stream $i$ and $\sigma$ is the index of currently served traffic stream. For more detailed information on the self-organised traffic control strategy see the paper by L"ammer and Helbing \cite{bplaczek:bib12}.
\selectlanguage{english}
\subsection{Traffic model}
Traffic model in the traffic control system is used to estimate the numbers of vehicles approaching an intersection ($N_i$) and to predict the required green times ($G_i$). In contrast to the original self-organised control method, which uses a macroscopic (fluid dynamic) traffic model \cite{bplaczek:bib12}, the proposed approach is based on the microscopic fuzzy cellular model \cite{bplaczek:bib13}. This modification enables a better utilisation of the data acquired from VSN, concerning the parameters of particular vehicles.

The fuzzy cellular traffic model was formulated as a hybrid system combining cellular automata and fuzzy calculus. It was based on a cellular automata approach to traffic modelling that ensures the accurate simulation of real traffic phenomena \cite{bplaczek:bib14}. A characteristic feature of this model is that it uses fuzzy numbers to represent vehicles positions, velocities and other parameters. Moreover, the model transition from one time step to the next is based on arithmetic of the ordered fuzzy numbers. This approach benefits from advantages of the cellular automata models and eliminates their main drawbacks i.e. necessity of multiple Monte Carlo simulations and calibration issues \cite{bplaczek:bib15}.

A traffic lane in the fuzzy cellular model is divided into cells that correspond to the road segments of equal length. Road traffic streams at an intersection are represented as sets of vehicles. A vehicle $j$ in traffic lane $i$ is described by its position $X_{i,j}$ (occupied cell) and velocity $V_{i,j}$ (in cells per time step). The maximum velocity is defined by the parameter $V_{max}$. The velocities and positions of all vehicles are updated simultaneously in discrete time steps of one second. All the above mentioned variables are expressed by fuzzy numbers.

In this study it was assumed that the fuzzy numbers have trapezoidal or triangular membership functions, thus they are represented by four scalars and the notation $A=(a^{(1)},a^{(2)},a^{(3)},a^{(4)})$  is used. Arithmetic operations are computed for the fuzzy numbers using the following definition:
\begin{equation}
   o(A,B)=(o(a^{(1)},b^{(1)}),o(a^{(2)},b^{(2)}),o(a^{(3)},b^{(3)}),o(a^{(4)},b^{(4)})),
\end{equation}
where $A$, $B$ are the fuzzy numbers and $o$ stands for an arbitrary binary operation.

The application of fuzzy calculus helps to deal with incomplete traffic data and enables straightforward determination of the uncertainty in simulation results \cite{bplaczek:bib16}. The main advantage of the fuzzy cellular model relies on the fact that the prediction of the parameters $N_i$ and $G_i$ is computationally efficient and the results are also represented by means of fuzzy numbers, thus their uncertainties can be easily evaluated.

\section{Data collection algorithms}
This section introduces three data collection algorithms for VSNs as well as defines the uncertainty estimates that enable reduction of data transmission. The data collection algorithms are presented as components of the traffic control procedure, which was discussed in Section 3. It should be also noted here that the control procedure is executed independently for each intersection in the road network. 

At first, some basic operations will be explained, which are common to the three proposed algorithms (see the pseudocodes below). The aim of the model \emph{update} operation is to approximate the current state of the traffic flow i.e. current positions of all vehicles approaching an intersection ($X_{i,j}$). This approximation is based on both the real traffic data acquired from VSN and the results of real-time simulation. During the real-time simulation the traffic model is used to estimate the missing positions of vehicles that were excluded from direct data acquisition. Besides the data on vehicle positions, the model update operation has to take into account also the real-time status data of traffic control operations (i.e. current traffic signals).

As it was mentioned in the previous section, the traffic model is used to \emph{predict} the numbers of vehicles approaching an intersection ($N_i$) and the required green times ($G_i$). Values of these parameters for all lanes are predicted by faster than real-time simulation using the approximation of current traffic state to determine initial conditions. The prediction results are used to \emph{make control decision} that determines which traffic stream should get a green signal at an intersection. Finally, traffic control system \emph{executes the control decision} by switching the appropriate signals.
%
%
\begin{figure}
\begin{Verbatim}[commandchars=\\\{\}] 
	for each time step
		update traffic model
		for each lane i=1..m
			for each vehicle j=1..n(i)
				if unc(X\textsubscript{i,j})>ut\textsubscript{pos} then acquire X\textsubscript{i,j}
			if new data collected then update traffic model
			for each lane i=1..m predict N\textsubscript{i}, G\textsubscript{i}
		make control decision
		execute control decision
\end{Verbatim} 
\caption{Pseudocode of data collection algorithm 1}
\end{figure}
Figure 2 shows pseudocode of the first data collection algorithm. According to this algorithm the vehicle position is acquired from VSN only if uncertainty of the position $unc(X_{i,j})$, approximated by traffic model, is higher than a predetermined threshold $ut_{pos}$ (in cells). This approach is similar to the concept of model-based querying, which was mentioned in Section 2. Note that the vehicle position is represented by a fuzzy number $X_{i,j}=(x^{(1)}_{i,j},x^{(2)}_{i,j},x^{(3)}_{i,j},x^{(4)}_{i,j})$. In order to estimate its uncertainty, the definition was adapted, which is based on determination of the area under membership function. Using this definition, the following formula was derived:
\begin{equation}
   unc(X_{i,j})=0.5|x^{(1)}_{i,j}-x^{(2)}_{i,j}|+|x^{(2)}_{i,j}-x^{(3)}_{i,j}|+0.5|x^{(3)}_{i,j}-x^{(4)}_{i,j}|
\end{equation}

The second data collection algorithm (Fig. 3) estimates uncertainty of the predicted green times $unc(G_i)$ using a similar measure to that defined in (5).  If for a given lane ($i$) the prediction uncertainty of $G_i$ is higher than a threshold value $ut_{pred}$ (in seconds) then the positions of vehicles in that lane are acquired. During the data acquisition only those vehicles are taken into account whose positions cannot be precisely determined by the traffic model. 

%
\begin{figure}
\begin{Verbatim}[commandchars=\\\{\}] 
	for each time step
		update traffic model
		for each lane i=1..m predict N\textsubscript{i}, G\textsubscript{i}
			if unc(G\textsubscript{i})>ut\textsubscript{pred} then
				for each vehicle j=1..n(i)
					if unc(X\textsubscript{i,j})>0 then acquire X\textsubscript{i,j}
		if new data collected then
			update traffic model
			for each lane i=1..m predict N\textsubscript{i}, G\textsubscript{i}
		make control decision
		execute control decision
\end{Verbatim} 
\caption{Pseudocode of data collection algorithm 2}
\end{figure}

Uncertainty of control decisions is used for detecting the necessity of data transfers in the third data collection algorithm (Fig. 4). The decision uncertainty is estimated as the maximum of uncertainties associated with the two rules of the control strategy i.e. the stabilisation and the optimisation rule: $unc(decision)=\max(unc_{stab}, unc_{opt})$.

The decision rules include comparison operations that have to be executed for fuzzy numbers and thus the probabilistic approach to fuzzy numbers comparison \cite{bplaczek:bib17} is employed. This approach enables estimation of the probability $P$ with which one fuzzy number is less, greater or equal to another fuzzy number. The probabilities are used for uncertainty estimation of traffic control decisions. Detailed discussion of the decision uncertainty estimation method can be found in \cite{bplaczek:bib11}, here only the resulting formulas are given with short comments.

%
\begin{figure}
\begin{Verbatim}[commandchars=\\\{\}] 
	for each time step
		update traffic model
		for each lane i=1..m predict N\textsubscript{i}, G\textsubscript{i}
		make control decision
		if unc(decision)>ut\textsubscript{dec} then 
			for each lane i=1..m
				for each vehicle j=1..n(i)
					if unc(X\textsubscript{i,j})>0 then acquire X\textsubscript{i,j}
		if new data collected then
			update traffic model
			for each lane i=1..m predict N\textsubscript{i}, G\textsubscript{i}
			make control decision
		if unc(decision)<=ut\textsubscript{dec} then execute control decision
\end{Verbatim} 
\caption{Pseudocode of data collection algorithm 3}
\end{figure}

Let $\sigma$ denote the result of the control decision i.e. an index of traffic stream which will get a green signal. The stabilisation rule determines $\sigma$ when $Z_{\sigma} \geq T_{max}$. It was assumed for this study that the condition $Z_{\sigma} \geq T_{max}$ is satisfied if the probability $P(Z_{\sigma} \geq T_{max})$ is above 0.5. Thus, in opposite situation the optimisation rule is activated. Such assumptions lead to the following definition of the stabilisation uncertainty:
\begin{equation}
   unc_{stab}=\left\{ 
  \begin{array}{l l}
     2P(Z_\sigma<T_{max}) & \quad \mathrm{if} \: P(Z_{\sigma} \geq T_{max})>0.5\\
     2P(Z_\sigma \geq T_{max}) & \quad \mathrm{if} \: P(Z_{\sigma} \geq T_{max})\leq0.5.\\
   \end{array} \right.
\end{equation}

The optimisation uncertainty corresponds with the comparisons that are necessary for finding the highest priority value $\pi_\sigma$. This uncertainty equals zero if the stabilisation rule determines the control decision: 
\begin{equation}
   unc_{opt}=\left\{ 
  \begin{array}{l l}
     0 & \quad \mathrm{if} \: P(Z_{\sigma} \geq T_{max})>0.5\\
     \max_i 2P(\pi_\sigma < \pi_i) & \quad \mathrm{if} \: P(Z_{\sigma} \geq T_{max})\leq0.5.\\
   \end{array} \right.
\end{equation}
The symbols used in (6) and (7) were defined in Section 3.1. Resulting values of the uncertainties $unc_{stab}$, $unc_{opt}$ and $unc(dec)$ are in range between 0 and 1. 
%
\section{Experimental results}
The proposed data collection algorithms were applied to the VSN-based traffic control in a road network. The experiments were performed in a traffic simulator which was developed for this purpose on the basis of Nagel-Schreckenberg stochastic cellular automata \cite{bplaczek:bib14}. Structure of the simulated network is presented in Fig. 5. Roads are unidirectional, thus each intersection has two incoming traffic streams. Links between intersections consists of 40 cells that correspond to the distance of 300 m. Maximal velocity of vehicles is 2 cells per time step i.e. 54 km/h (the simulation time step is one second).  Deceleration probability $p$ is 0.15. For the above settings of the Nagel-Schreckenberg traffic model, the obtained saturation flow at intersections is about 1700 vehicles per hour of green time. 

The self-organised traffic control was simulated assuming the intergreen times $\tau$ of 5 s and the maximum period $T_{max}$ of 120 s.
Architecture of the considered traffic control system consists of two types of VSN nodes: control and vehicle nodes. The fixed control nodes installed at intersections collect sensor data from the vehicle nodes and execute the traffic control procedure. Each vehicle in the system is equipped with a wireless communication unit and uses a GPS device to determine its position. Every time a vehicle enters the road network, it has to register itself by sending a hello message. The data collection operation is initialised by the control node which generates queries to acquire positions of vehicles approaching an intersection.
 %
%
\begin{figure}
\centering
\includegraphics [height=5cm] {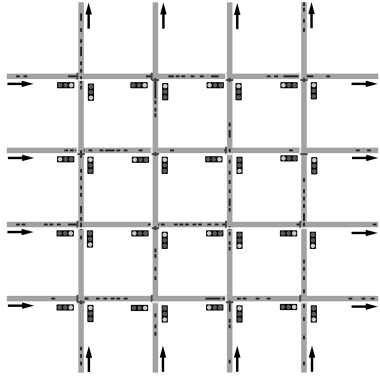}
\caption{Simulated road network}
\end{figure}

Simulation results for the three data collection algorithms are compared in Fig. 6. The comparison takes into account the performance of traffic control and the number of data transfers from particular vehicles to the control nodes. The average delays of vehicles and data transfer counts were determined from 3 hour traffic simulations. In this experiment, the traffic flow volumes were changed gradually in order to reproduce saturation levels (demand-capacity ratios) from 0 to 100\%. 

The highest accuracy of the traffic information in the control system was obtained using the first data collection algorithm with threshold value $ut_{pos}=0$. This scenario results in lowest delays and highest number of data transfers. The delays grow drastically for algorithm 1 if a threshold value above 0 is used (see the plot for $ut_{pos}=5$). In comparison, the second algorithm for $ut_{pred}=5$ provides low delays and reduces the data transfer counts. However, for algorithm 2 with higher threshold values an increase of the delays is observed especially at low saturation levels. The best results were obtained for algorithm 3, which enables significant reduction in the data transfers and does not decrease the performance of the traffic control.
%
%
\begin{figure}
\centering
\includegraphics [height=5cm] {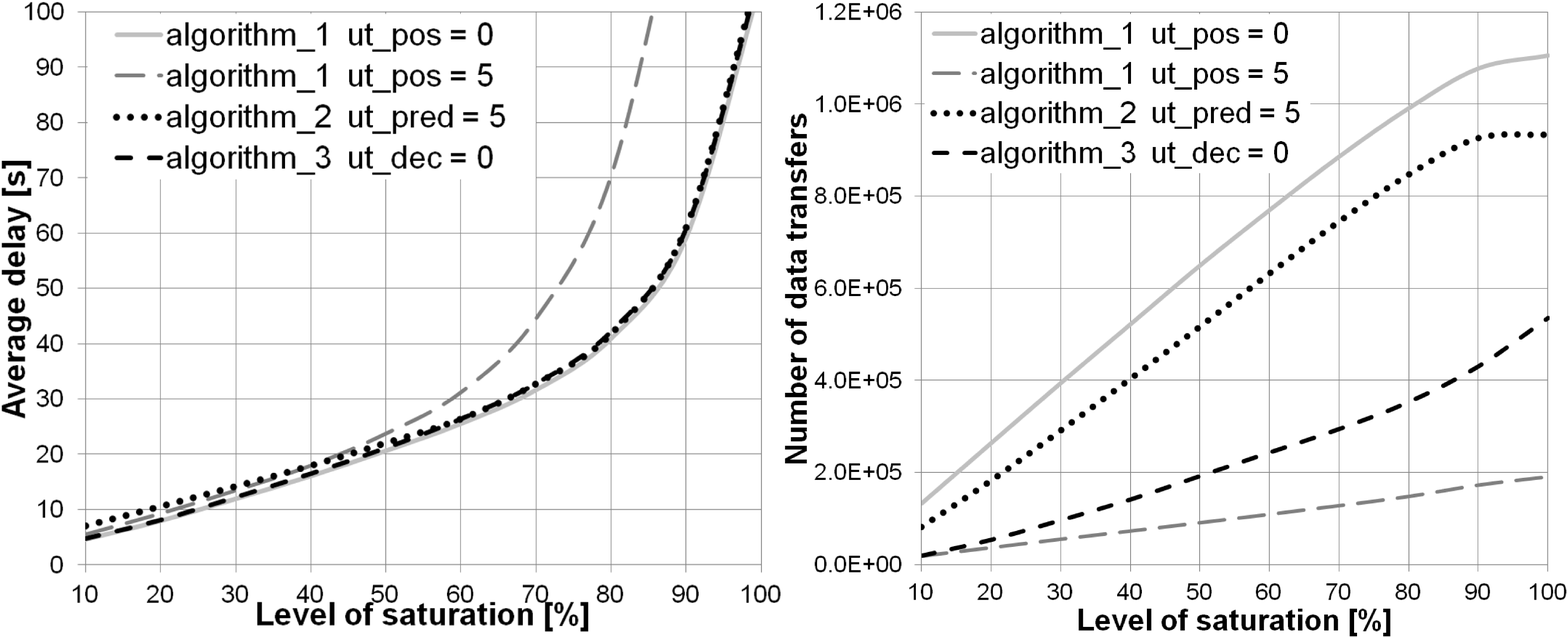}
\caption{Simulation results: average delay (left) and number of data transfers (right)}
\end{figure}
%

%
\section{Conclusion}
In this paper three data collection algorithms were proposed for VSN-based traffic control systems. Effectiveness of the introduced algorithms was evaluated in an experimental study on the traffic control in a road network. Experiments were carried out using a simulation environment. The tests confirmed that the proposed algorithms enable reduction in the data transmission for a wide range of traffic conditions. The most promising results were obtained for the algorithm using decision uncertainty to detect the necessity of data transfers.
%
%

\end{document}